\documentstyle[aps,prb,multicol,epsfig]{revtex}

\newcommand{\simle}
{\raisebox{-0.75ex}[-1.5ex]{$\;\stackrel{<}{\sim}\;$}}
\newcommand{\simge}
{\raisebox{-0.75ex}[-1.5ex]{$\;\stackrel{>}{\sim}\;$}}
\def\d{\partial}
\def\s{{\sigma}}
\def\e{{\epsilon}}
\def\k{{ {\bf k} }}

\def\q{{ {\bf q} }}

\def\w{{\omega}}
\def\a{{\alpha}}

\def\i{{ {\rm i} }}

\def\I{{ \mbox{\scriptsize I} }}
\def\II{{ \mbox{\scriptsize II} }}

\begin{document}
\draft

\def\runtitle{
Theory of the Hall Coefficient and the Resistivity 
on the Layered Organic Superconductors $\kappa$-(BEDT-TTF)$_2$X
}
\def\runauthor
 {Hiroshi {\sc Kontani} and Hiori {\sc Kino}}

\title{
Theory of the Hall Coefficient and the Resistivity \\
on the Layered Organic Superconductors $\kappa$-(BEDT-TTF)$_2$X
}

\author{
Hiroshi {\sc Kontani}$^{1,2}$
 \thanks{present address: Department of Physics, Saitama University,
 255 Shimo-Okubo, Urawa-city, 338-8570, Japan.}
 and Hiori {\sc Kino}$^{2}$
}

\address{
$^1$Theoretische Physik III,
 Universitaet Augsburg
 D-86135 Augsburg, Germany.
\\
$^2$Institute for Solid State Physics, University of Tokyo,
Kashiwanoha, Kashiwa-shi, 277-8581, Japan.
}

\date{\today}

\maketitle      

\begin{abstract}
In the organic superconducting $\kappa$-(BEDT-TTF)$_2$X compounds,
various transport phenomena exhibit striking
non-Fermi liquid behaviors, 
which should be the important clues to understanding
the electronic state of this system.
Especially, the Hall coefficient ($R_{\rm H}$) shows Curie-Weiss
type temperature dependence, which is similar to that of
high-$T_{\rm c}$ cuprates.
In this paper, 
we study a Hubbard model on an anisotropic
triangular lattice at half filling,
which is an effective model of $\kappa$-(BEDT-TTF)$_2$X compounds.
Based on the fluctuation-exchange (FLEX) approximation,
we calculate the resistivity ($\rho$) and $R_{\rm H}$ 
by taking  account of the vertex corrections for the current,
which is necessary for satisfying the conservation laws.
Our theoretical results $R_{\rm H}\propto T^{-1}$
and ${\rm cot}\theta_{\rm H}\propto T^2$
explain the experimental behaviors well,
which are unable to be reproduced by the conventional Boltzmann 
transport approximation.
Moreover, we extend the standard Eliashberg's transport theory
and derive the more precise formula for the conductivity,
which becomes important at higher temperatures.
\end{abstract}


%

\begin{multicols}{2}
\narrowtext

\section{Introduction}
It is well-known that the superconducting organic compound
$\kappa$-(BEDT-TTF)$_2$X systems
exhibit rich variety of ground states,
through the strong correlation effects between electrons
 \cite{Kanoda-review}.
For example, 
X=Cu[N(CN)$_2$]Cl salt is in the antiferromagnetic (AF) 
insulating phase in a low pressure region, $P<200$bar.
With increasing the pressure, it changes to the 
superconducting (SC) phase at the transition temperature 
$T_{\rm c}=13$K through the weak first-order transition,
and the SC phase disappears under $P\simge 10$kbar.
Above $T_{\rm c}$, $1/T_1T \propto T^{-1}$ is observed 
for a wider range of temperatures, which reflects 
the growth of the AF fluctuations as the temperature decreases
 \cite{relaxation}.
Thus, it is natural to consider that the AF fluctuation 
is the origin of superconductivity.

Recently, several theoretical works on 
$\kappa$-(BEDT-TTF)$_2$X system
were done by using the fluctuation-exchange (FLEX) approximation,
which is a kind of self-consistent spin-fluctuation theory.
They showed that the $d$-wave like superconductivity is induced
by the strong AF fluctuations
 \cite{Kino,Kondo,Schmalian,Kondo2}.
Moreover, characteristic features of the experimental 
pressure-temperature phase diagram were reproduced well
 \cite{Kino,Kondo2}.

In this system, various transport phenomena above $T_{\rm c}$
also show interesting non-Fermi liquid behaviors.
Recently, the temperature dependence of the resistivity ($\rho$)
and the Hall coefficient ($R_{\rm H}$) for X=Cu[N(CN)$_2$]Cl
were measured precisely above $T_{\rm c}$ under $P= 4.5\sim10$kbar
in Ref.
 \cite{Sushko}.
According to the measurement,
the approximate relations 
$\rho \propto T$ and $R_{\rm H} \propto T^{-1}$ 
are observed for $T=30\sim100$K, and
the Hall angle ${\rm cot}\theta_{\rm H}\equiv(\s_{xx}/{\mit\Delta}\s_{xy})$
is proportional to $T^2$ well.
In other measurement on X=Cu[NCS]$_2$, 
$R_{\rm H}$ increases by a factor of three 
on cooling below 60K at ambient pressure ($T_{\rm c}=10$K)
 \cite{Keizo}.
The mechanism of these interesting non-Fermi liquid behaviors, 
which are also observed in high-$T_{\rm c}$ cuprates, 
should be understood consistently.

In this paper, we present the 
theoretical study on both $\rho$ and $R_{\rm H}$
for $\kappa$-(BEDT-TTF)$_2$X
by using the FLEX approximation.
Based on the conserving approximation,
all the vertex corrections (VC's) for the current 
which is necessary to satisfy conserving laws are taken into account.
We find that the Curie-Weiss like behavior of $R_{\rm H}$ is naturally
reproduced by the VC's
when the AF fluctuations are dominant.
On the other hand, the conventional Boltzmann approximation,
which does not include any VC's,
fails to reproduce the temperature dependence of $R_{\rm H}$.
Experimentally,
an intimate relation between the AF fluctuations
ans the transport phenomena is recognized 
\cite{Sushko}.

Note that the effect of VC's in nearly AF Fermi liquid 
was first studied in high-$T_{\rm c}$ cuprates by refs.
 \cite{Kontani-Hall} and
 \cite{Kanki-Hall},
and the overall behavior of $R_{\rm H}$ are naturally 
reproduced both for 
hole-doped compounds and for electron-doped compounds.
The present study is based on them basically.

\section{Electronic States given by the FLEX Approximation}
\begin{figure}
\vspace{10mm}
\begin{center}
\epsfig{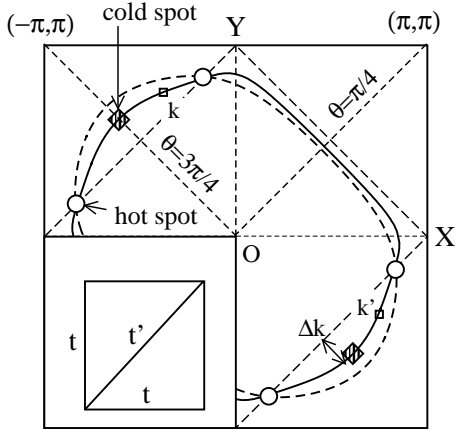}
\end{center}
\vspace{15mm}
\caption{The Fermi surface of the present model ($t'/t=0.7$)
 for $U=9$ (full line) and $U=0$ (broken line)
 determined by ${\rm Re}\{G_\k^{-1}(0)\}=0$.
 The long-dashed lines, connecting between $X$ and $Y$
 for example, represent the AF-zone boundary. 
 The hot spots locate on the AF-zone boundary.
 Inset: the lattice structure.}
\label{fig:FS}
\end{figure}

We study the triangular lattice Hubbard model with anisotropic hopping
parameters ($t$, $t'$) as shown in the inset of Fig. \ref{fig:FS},
which is a simple effective model for $\kappa$-(BEDT-TTF)$_2$X system
 \cite{HF-approx}.
The dispersion is given by 
\begin{eqnarray}
\e_\k^0= 2t(\cos(k_x)+\cos(k_y)) + 2t'\cos(k_x+k_y),
 \label{eqn:disp1}
\end{eqnarray}
where we put the lattice spacing 1.
We analyze this model by using the FLEX method, which is 
a kind of self-consistent perturbation theory.
This method had been applied to the study of high-$T_{\rm c}$ cuprates,
and various non-Fermi liquid behaviors were reproduced well
 \cite{FLEX_QMC,2D-SC-Monthoux,high-Tc-Dahm}.
It has also been applied to the superconducting ladder compound, 
Sr$_{14-x}$Ca$_x$Cu$_{24}$O$_{41}$
 \cite{Trellis}.
 
In the present study we put $t=1$ and $t'=0.7$,
where a $d$-wave like superconductivity
is realized at $T_{\rm c}>0.02$ for $U\ge7$,
and it is replaced by the AF phase for $U\ge10$
 \cite{Kino}.
Figure \ref{fig:FS}
shows the Fermi surfaces at half-filling 
for $U=0$ and $U=9.0$ at $T=0.02$.
The Fermi surface is hole-like because $t,t'>0$.
There are two reflection symmetries with respect to 
the $(\theta\!=\!\pi/4)$-axis and the $(\theta\!=\!3\pi/4)$-axis
in Fig. \ref{fig:FS}.
We see that 
the nesting of the Fermi surface is strengthened by 
the deformation of the Fermi surface in the case of finite $U$,
which is caused by the real part of the self-energy.

The self-energy in the FLEX approximation
is given by
\begin{eqnarray}
& &\Sigma_\k(\e_n) 
 = T\sum_{\q,l} G_{\k-\q}(\e_n-\w_l)\cdot V_\q(\w_l), 
 \label{eqn:self}\\
& &V_\q(\w_l)= U^2 \left( \frac32 {\chi}_\q^{s}(\w_l) 
  +\frac12 {\chi}_\q^{c}(\w_l) 
  - {\chi}_\q^0(\w_l) \right) \mbox{,}
    \\
& &{\chi}_\q^{s(c)}(\w_l)
 = {\chi}_\q^0(\w_l) \cdot \left\{ {1} -(+)
  U{\chi}_\q^0(\w_l) \right\}^{-1} \mbox{,} 
     \label{eqn:chi} \\
& &\chi_\q^0(\w_l)
 = -T\sum_{\k, n} G_{\q+\k}(\w_l+\e_n) G_\k(\e_n) \mbox{,}
     \label{eqn:chi0}
\end{eqnarray}
where $\e_n= (2n+1)\i\pi T$ and $\w_l= 2l\cdot\i\pi T$, respectively.
By noticing the Dyson equation
$\{ G_\k(\e_n) \}^{-1} = \e_n + \mu - \e_\k^0 - {\Sigma}_\k(\e_n)$,
we solve the eqs. (\ref{eqn:self})-(\ref{eqn:chi0}) self-consistently,
choosing the chemical potential $\mu$ 
so as to keep the system at half-filling.
Here we use 4096 $k$-meshes and $256$-Matsubara frequencies,
respectively.

\begin{figure}
\begin{center}
\epsfig{file=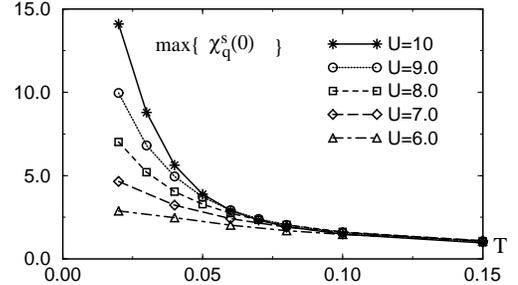,width=7.5cm}
\end{center}
\caption{The maximum value of $\chi_\q^s(0) \ (\propto \xi_{\rm AF}^2)$
for $U=6\sim10$.
} 
 \label{fig:Kai}
\end{figure}

Here, ${\chi}_\q^{s}(0)$ gives the static spin susceptibility.
Figure \ref{fig:Kai} shows the Curie-Weiss behavior of 
the maximum value of $\chi_\q^s(0)$,
which is proportional to the 
square of the AF-correlation length $\xi_{\rm AF}$.
The obtained relation $\xi_{\rm AF}^2\propto T^{-1}$ 
is known to be caused by the renormalization of the self-energy.
The FLEX approximation also gives the relation
$(T_1T)^{-1}=\sum_\k{\rm Im}\chi_\k^s(\w)/\w \propto T^{-1}$. 

Nonetheless $\chi_\q^s(0)$ for $U=0$ is incommensurate,
it becomes commensurate in the case of $U\ge8$ at $T=0.02$,
which is consistent with experiments.
This change of the shape of $\chi_\q^s(0)$
is brought by the deformation of the interacting Fermi surface
as shown in Fig. \ref{fig:FS},
which can not be reproduced by the simple renormalization of $t'/t$
  \cite{Kino}.
We also note that the obtained $\chi_\q^s(0)$ 
will be slightly overestimated at low temperatures 
because its VC's are neglected here.

\begin{figure}
\vspace{-10mm}
\begin{center}
\epsfig{file=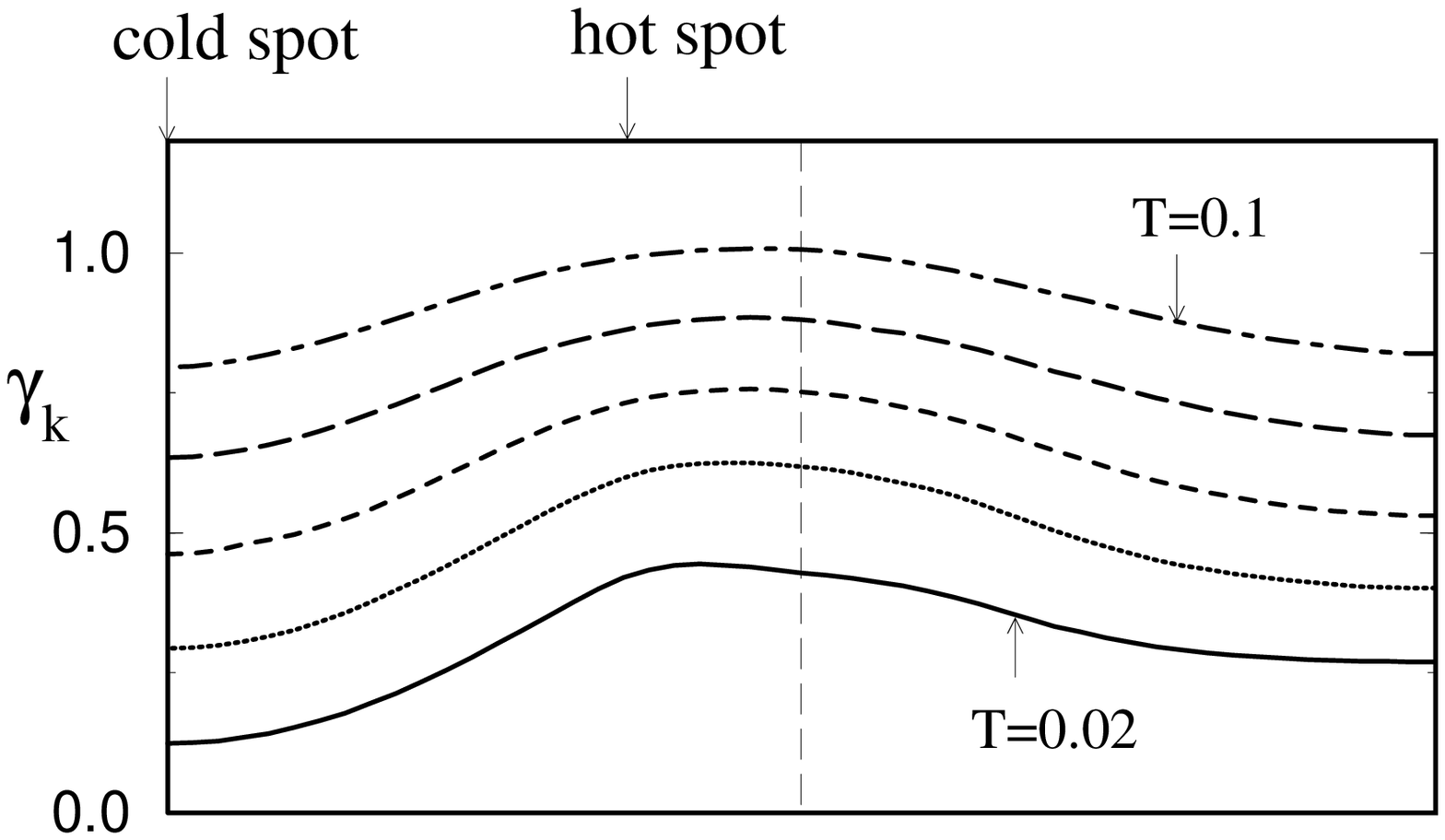,width=7.5cm}
\end{center}
\vspace{-35mm}
\begin{center}
\epsfig{file=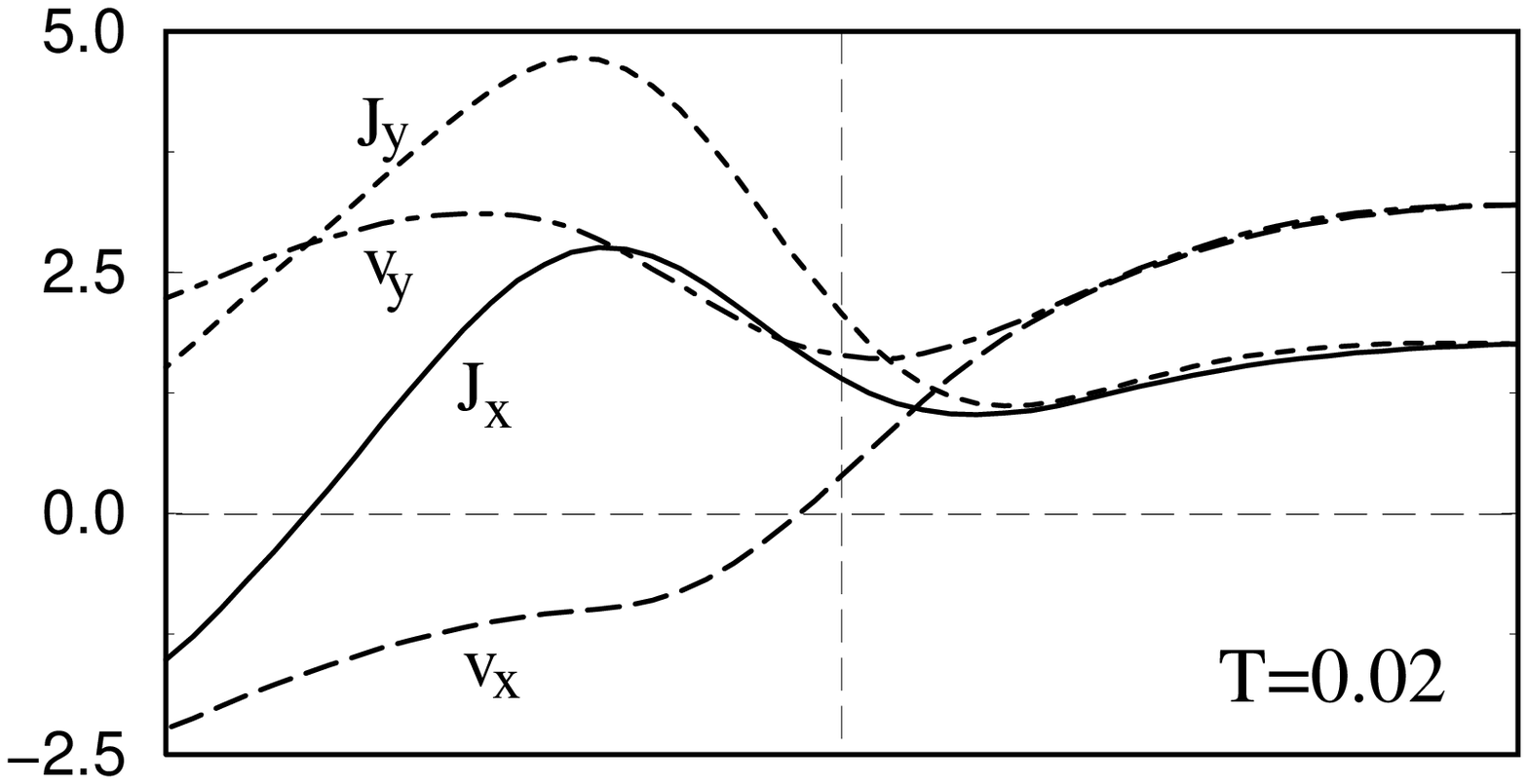,width=7.5cm}
\end{center}
\vspace{-35mm}
\begin{center}
\epsfig{file=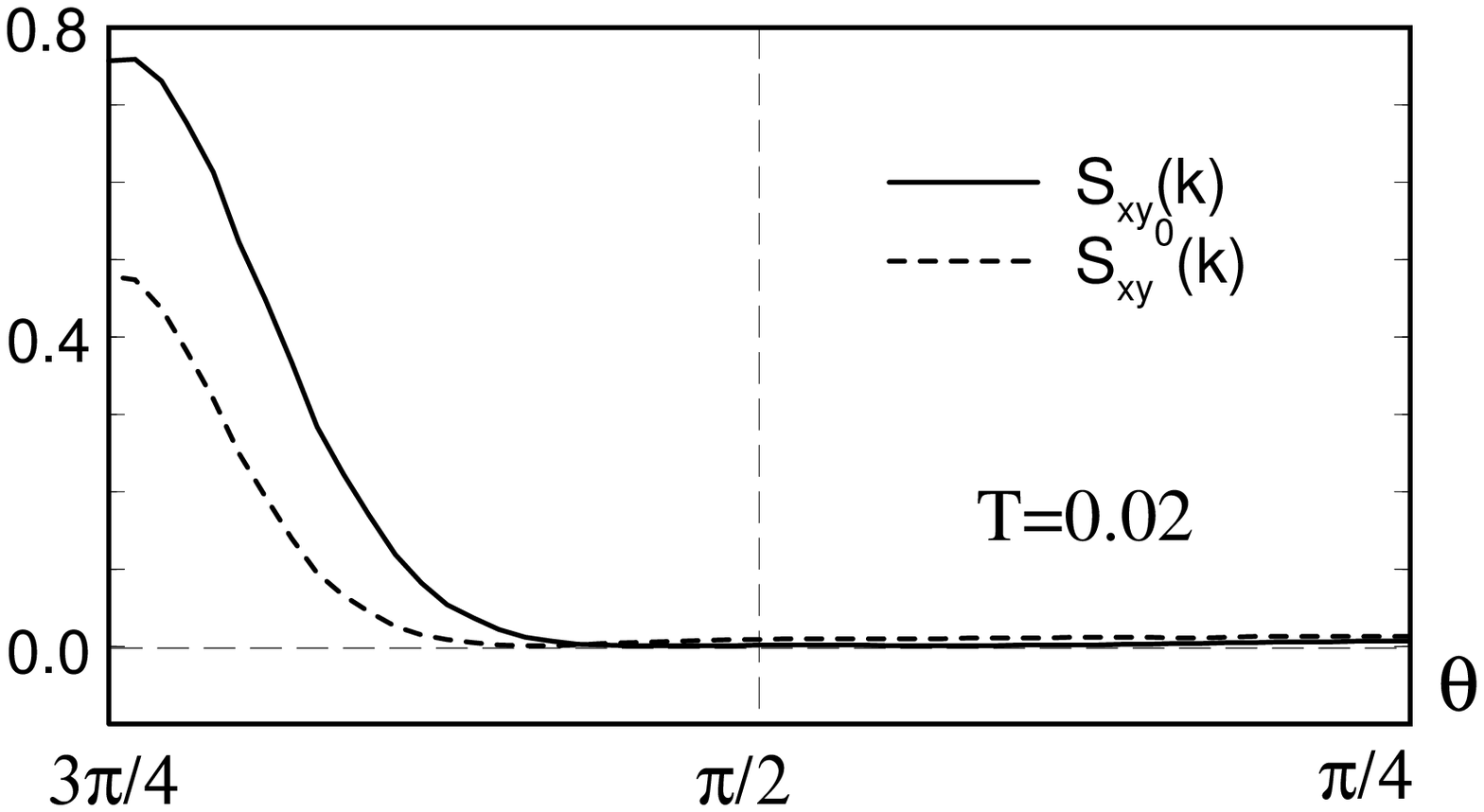,width=7.5cm}
\end{center}
\caption{(a) $\gamma_\k={\rm Im}\Sigma_\k(-\i0)$, 
 (b) ${\vec J}(\k,0)$ and ${\vec v}(\k,0)$, 
 (c) $S_{xy}(\k)$ and $S_{xy}^0(\k)$,
 along the Fermi surface for the region $\pi/4\le\theta\le3\pi/4$,
 for $U=9$.
 The locations of the 'hot spot' and the 'cold spot'
 are introduced in Fig. 1.
}
\label{fig:J}
\end{figure}

Next, Fig. \ref{fig:J} (a) shows the 
imaginary part of the self-energy, 
$\gamma_\k= {\rm Im}\Sigma_\k(-\i 0) >0$,
along the Fermi surface for the region $\pi/4\le\theta\le3\pi/4$
for $T=0.02,0.04,\cdots,0.1$.
The definition of the hot spots and the cold spots
are given in Fig. \ref{fig:FS}.
In this temperature region,
$\gamma_\k$ takes the minimum (maximum) value
at the the cold spot (hot spot).
A hot spot is separated from its counterpart
by ${\bf Q}=(\pi,\pi)$ in the reciprocal space,
and a cold spot is the most distant point from the AF-zone boundary.
In the present study,
the relations $\gamma_{\rm cold}\propto T$ and 
$\gamma_{\rm hot}\propto \sqrt{T}$ are satisfied because 
$\xi_{\rm AF} \simle {\mit\Delta}k^{-1}$
in the FLEX approximation, where ${\mit\Delta}k$
is explained in Fig. \ref{fig:FS}
 \cite{Kontani-Hall}.
Note that the damping rate of the quasiparticle is given by
$\gamma_\k^\ast = z_\k\gamma_\k$, where
$z_\k \equiv (1-\frac{\d}{\d\w}{\rm Re}\Sigma_\k(\w))_{\w=0}^{-1}$
is the renormalization factor.
In the numerical calculation for $U=9$, $\gamma_\k^\ast \simle T$ 
around the cold spots at lower temperatures
because $z_\k^{-1} \simle 10$ there.
Thus, quasiparticle can still be defined.

\section{Hall Conductivity Based on the Conservation Approximation}
\subsection{Derivation of the Total Current ${\vec J}_{\k}(\w)$
Based on the Conservation Approximation}
In this section, we calculate both $\s_{xx}$ and ${\mit\Delta}\s_{xy}$
based on the conserving approximation.
By using the Kubo formula, they are derived as
 \cite{Eliashberg,Kohno-Yamada,Fukuyama}
\begin{eqnarray}
& &\sigma_{xx} = {e}^2 \sum_\k \int_{-\infty}^\infty \frac{d\e}{\pi}
 \left(-\frac{\d f}{\d\e} \right) \left( \
   |G_\k(\e)|^2 \cdot v_{\k x}(\e) J_{\k x}(\e) \right.
 \nonumber \\
& &\ \ \ \ \ \ \ \ \ \ \ \
 - \left. {\rm Re}\{G_\k^2(\e) \cdot v_{\k x}^2(\e)\} \ \right),
  \label{eqn:s_numerical} \\
& &{\mit\Delta}\sigma_{xy} 
 = -B\cdot{e}^3 \sum_\k \int_{-\infty}^\infty \frac{d\e}{2\pi}
 \left(-\frac{\d f}{\d\e} \right) 
 S_{xy}(\k,\e), 
  \label{eqn:sH_numerical} \\
& &\ S_{xy}(\k,\e) =
  |G_\k(\e)|^2 \cdot |{\rm Im}G_\k(\e)|  \nonumber \\
& &\ \ \ \times 
 v_{\k x}(\e)\left[ J_{\k x}(\e) \frac{\d J_{\k y}(\e)}{\d k_y} 
     -J_{\k y}(\e) \frac{\d J_{\k x}(\e)}{\d k_y} \right] 
 \nonumber \\
& & \ \ \  + \ \langle x \leftrightarrow y \rangle,  
  \label{eqn:A_numerical}
\end{eqnarray}
where $f(\e)=(\exp((\e-\mu)/T)+1)^{-1}$, and
$G_\k(\w+\i\delta)$ and $\Sigma_\k(\w+\i\delta)$ are derived 
from $G_\k(\w_n)$ and $\Sigma_\k(\w_n)$
through the numerical analytic continuation.
$B$ is the magnetic field parallel to the $z$-axis.
$v_{\k \mu}(\w) = \frac{\d}{\d k_\mu} 
 \left( \e_\k^0 + {\rm Re}\Sigma_\k(\w) \right)$
is the quasiparticle velocity,
and $J_{\k \mu}(\w)$ is the total current
which contains the vertex correction from ${\cal T}_{22}$
in the notation of Ref.
 \cite{Eliashberg}.
Later, we examine its importance in detail.

As for the resistivity,
the second term of eq. (\ref{eqn:s_numerical})
is neglected in the Eliashberg's transport theory,
whose derivation is given in \S III C.
We call it the incoherent part of the conductivity, $\s_{\rm inc}$,
because it is negligible in the case of $\gamma_\k^\ast \ll T$.
As a result, these formulae 
(\ref{eqn:s_numerical})-(\ref{eqn:J_numerical})
are valid even for $\gamma_\k^\ast \sim T$.

\begin{figure}
\vspace{-10mm}
\begin{center}
\epsfig{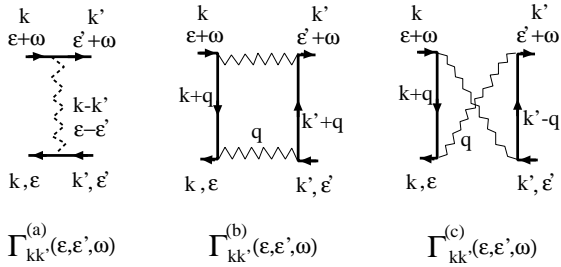}
\end{center}
\caption{All the irreducible VC's 
in the framework of the FLEX approximation:
(a) MT-term, and (b,c) AL-terms, respectively.
}
\label{fig:MTAL}
\end{figure}

In the present study, 
we solve the following Bethe-Salpeter equations 
for $J_{\k \mu}(\e)$
after the manner of the conserving approximation:
\begin{eqnarray}
& &\ J_{\k \mu}(\w) = v_{\k \mu}(\w)+ \sum_{\q} 
 \int_{-\infty}^\infty \frac{d\e}{2\pi}
 \left[ {\rm cth}\frac{\e-\w}{2T} - {\rm th}\frac{\e}{2T} \right] 
 \nonumber \\
& &\ \ \ \ \ \ \times 
 {\rm Im}V_{\k-\q}(\e-\w+\i\delta) \cdot 
 |G_\k(\e)|^2 \cdot J_{\q \mu}(\e) ,
  \label{eqn:J_numerical} 
\end{eqnarray}
where we take only the Maki-Thompson (MT) type VC's
into account, and neglect the Aslamazov-Larkin (AL) terms
because it has little contributions 
if $\chi_\q(0)$ has a sharp peak around ${\bf q}=(\pi,\pi)$
 \cite{Kontani-Hall};
see Fig. \ref{fig:MTAL}.
Figure \ref{fig:J} (b) shows the numerical solution
of eq. (\ref{eqn:J_numerical}).
We see that 
${\vec J}_\k$ is not parallel to ${\vec v}_\k$.

On the other hand, in the Boltzmann theory
within the relaxation time approximation (RTA), 
$J_{\k \mu}(\e)$ is simply replaced by $v_{\k \mu}(\e)$ in eqs.
(\ref{eqn:s_numerical})-(\ref{eqn:A_numerical}).
It is an insufficient approximation for the Hall coefficient
in nearly AF state as shown in Refs.
 \cite{Kontani-Hall,Kanki-Hall}.

Now we discuss
why ${\vec J}_\k$ shows such an anomalous behavior.
Here, we choose an arbitrary point on the Fermi surface, $\k$,
which locates between two hot spots in the region $\pi/2<\theta<\pi$,
and also define that $\k'\equiv (k_y,k_x)$.
(see Fig. \ref{fig:FS}.)
Then, $\k\!-\!\k' \!\approx\! (-\pi,\pi)$ is satisfied
in the present system, as shown in Fig. \ref{fig:FS}.
In the Bethe-Salpeter eq. (\ref{eqn:J_numerical}),
${\vec J}_\k$ are strongly connected with  ${\vec J}_{\k'}$
through $V_{\k-\k'}(\w)$ 
in the presence of strong AF fluctuations.
Taking this fact into account,  
the approximate solution of eq. (\ref{eqn:J_numerical}) is given by
 \cite{Kontani-Hall}
\begin{eqnarray}
{\vec J}_\k = ({\vec v}_\k+\a_\k{\vec v}_{\k'})\cdot(1-\a_\k^2)^{-1},
 \label{eqn:J_analytical}
\end{eqnarray}
where $\a_\k = \langle \cos(\theta_J(\k)-\theta_J(\q)) 
 \rangle_{|\q-\k|<\xi_{\rm AF}^{-1}}$,
and $\theta_J(\q)$ is the angle of ${\vec J}_\q$. 
According to the definition,  
$\a_\k<1$ and $(1-\a_\k)^{-1} \propto \xi_{\rm AF}^{2}$
 \cite{Kontani-Hall}.
Equation (\ref{eqn:J_analytical}) means that
$J_\k$ at the hot spots will become parallel to the 
AF zone boundary when $\xi_{\rm AF}$ approaches to infinity.
In fact,shch a tendency is recognized in Fig. \ref{fig:J} (b).
Thus, the anomalous behavior of ${\vec J}_\k$,
which is the result of the multiple scattering 
between ${\vec v}_\k$ and ${\vec v}_{\k'}$, 
becomes quite singular when the AF fluctuations are dominant.
As a result, the RTA 
is strongly violated when $\xi_{\rm AF}\gg1$
 \cite{SDW-comment}.


Equation (\ref{eqn:sH_numerical}) is rewritten at low temperatures as
\begin{eqnarray}
& &{\mit\Delta}\s_{xy}
 = B\cdot\frac{e^3}{4}\int_{\rm FS} dk_\parallel S_{xy}(k_\parallel)
 \nonumber \\
& &\ \ \ \ S_{xy}(k_\parallel)
 = ( {\vec J}_{\k} \times 
  d{\vec J}_{\k}/d k_\parallel )_z /\gamma_{\k}^2
\nonumber \\
& &\ \ \ \ \ \ \ \ \ \ \ \ \ \ \ 
 = |{\vec J_{\k}}|^2 
 (d \theta_J({\k})/d k_\parallel) /\gamma_{\k}^2,
 \label{eqn:simple}
\end{eqnarray}
where 
$\int_{\rm FS} dk_\parallel$ represents the 
momentum integration along the Fermi surface, and 
$dk_\parallel$ is parallel to the Fermi surface
 \cite{Kontani-Hall,Kanki-Hall}.

Because $d\a_\k/dk_{\parallel}=0$ at the cold spot
due to the symmetry of this model,
$( {\vec J}_\k\times d{\vec J}_\k/d k_\parallel )_z
 = (1-\a_\k^2)^{-1} \cdot
 ( {\vec v}_\k\times d{\vec v}_\k/d k_\parallel )_z$
at the cold spot by using eq. (\ref{eqn:J_analytical}).
Because ${\mit\Delta}\s_{xy}$ is given mainly around the cold spots
in the $k_\parallel$-integration of eq. (\ref{eqn:simple}),
we conclude the relation
\begin{eqnarray}
{\mit\Delta}\s_{xy}/{\mit\Delta}\s_{xy}^0 
 \ \propto \xi_{\rm AF}^2 \ \propto T^{-1}
 \label{eqn:sxysxy0}
\end{eqnarray}
in the presence of the strong AF fluctuations.
Here ${\mit\Delta}\s_{xy}^0$ is given by 
replacing ${\vec J}_\k(\w)$ with ${\vec v}_\k(\w)$ 
in eq.(\ref{eqn:sH_numerical}),
which is equal to the result of the RTA.
As for ${\mit\Delta}\s_{xy}^0$,
the conventional Kohler's rule
${\mit\Delta}\s_{xy}^0 \propto (\s_{xx})^2$ is well satisfied
in the present calculation because 
the anisotropy of $\gamma_\k$ is not so extreme.
($\gamma_{\rm hot}/\gamma_{\rm cold}$ is at most 3.)
As a result, eq. (\ref{eqn:sxysxy0}) leads the relation
$R_{\rm H}\equiv({\mit\Delta}\s_{xy}/B)\cdot \rho^2 
 \propto \xi_{\rm AF}^2 \propto T^{-1}$, 
which is recognized 
in the present numerical calculations as shown below.

The above analysis is confirmed by 
the numerical results of $S_{xy}(k_\parallel)$ 
in Fig. \ref{fig:J} (c).
$S_{xy}^0(k_\parallel)$ is given by the
RTA.
The dominant contributions to ${\mit\Delta}\s_{xy}$ 
come from the region around the cold spot,
$S_{xy}(k_\parallel)$ for region $\pi/4<\theta<\pi/2$
is considerably small because the curvature of the
Fermi surface is very small there.
In the present case, both 
$S_{xy}^0(k_\parallel)$ and $S_{xy}(k_\parallel)$ 
are positive everywhere.
It is not the case for high-$T_{\rm c}$ cuprates;
for example, the change of the sign of $R_{\rm H}$ is realized 
in Nd-compounds
 \cite{Kontani-Hall,Kanki-Hall}.

\subsection{Numerical Results for Transport Phenomena}
Now, we study both $\rho$ and $R_{\rm H}$
for various values of $U$,
because the main effect of the applied pressure is expected to 
increases the bandwidth $W_{\rm b}$, in other words,
to reduce the value of $U/W_{\rm b}$
 \cite{HF-approx}.
At the same time, the value of $t'/t$ may be also modified 
by pressure.
This effect is not discussed here
although the electronic states are sensitive to  $t'/t$
according to the FLEX approximation
 \cite{Kino}.

\begin{figure}
\vspace{5mm}
\begin{center}
\epsfig{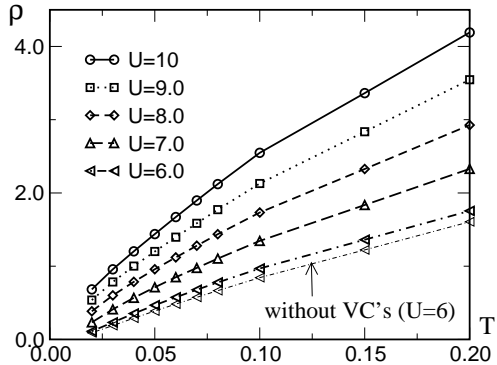}
\end{center}
\vspace{5mm}
\begin{center}
\epsfig{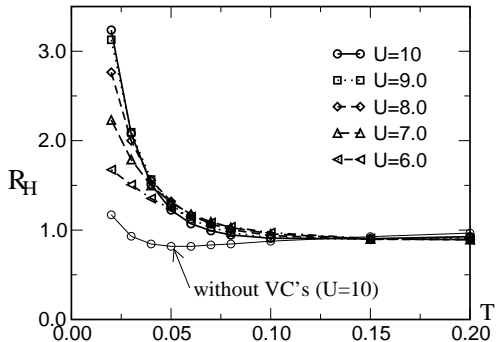}
\end{center}
\vspace{10mm}
\begin{center}
\epsfig{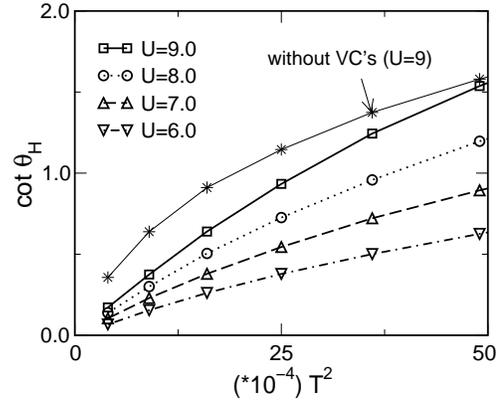}
\end{center}
\vspace{13mm}
\caption{The temperature dependence of (a) $\rho$, (b) $R_{\rm H}$,
 and (c) the cot$\theta_{\rm H}$ for $U=6\sim10$.
 We put $e=1$ and $\hbar=1$.
 $T=0.1$ corresponds to 100K approximately.
}
\label{fig:Result}
\end{figure}

Figure \ref{fig:Result} (a) shows the temperature 
dependences of the resistivity $\rho=1/\s_{xx}$.
All the $\rho$'s show the approximate $T$-linear resistivity,
reflecting the $T$-linear behavior of $\gamma_{\rm cold}$.
At the same temperature, 
$\rho$ increases monotonously as $U$ increases.
Here, $\rho_0=1/\s_{xx}^0$
is the resistivity without the VC's, which is given
by replacing ${\vec J}_\k(\w)$ with ${\vec v}_\k(\w)$ in eq.
(\ref{eqn:s_numerical}).
We see that 
$\rho > \rho_0$ due to the VC's for the current.

Here, we comment on the anisotropy of $\rho$.
In the present model,
$\s_{\mu\mu}$ depends on the angle of the $\mu$-axis
because there is no four-fold rotational symmetry in this system.
We find that $\rho$ takes its maximum (minimum) value 
along the ($\theta\!=\!\pi/4$)-axis (($\theta\!=\!3\pi/4$)-axis),
and $\rho_{\rm max}/\rho_{\rm min} \approx 1.6$
with weak temperature dependence.

Next, we discuss on the Hall coefficient $R_{\rm H}$.
As shown in Fig. \ref{fig:Result} (b),
$R_{\rm H}$ follows the Curie-Weiss like temperature dependence, 
which is consistent with the relation (\ref{eqn:sxysxy0}).
Actually, both $\xi_{\rm AF}^2\propto \chi_Q(0)$ and $R_{\rm H}$ 
increase monotonously as $U$ increases at the same temperatures.
(see Fig. \ref{fig:Kai}.)
As a result, the large temperature dependence of $R_{\rm H}$ 
in $\kappa$-(BEDT-TTF)$_2$X salts is reproduced well
by taking account of the VC's for the current.

To see the importance of the VC's,
we also show $R_{\rm H}^0=({\mit\Delta}\s_{xy}^0/B)\cdot \rho_0^2$, 
which is the result of the RTA:
We see that $R_{\rm H}^0$ is nearly constant,
and $R_{\rm H}\approx R_{\rm H}^0$  
at higher temperatures ($T\simge0.1$)
because of ${\vec J}_\k \approx {\vec v}_\k$ in this case.
We note that $dR_{\rm H}^0/dT$ is slightly positive for $T>0.05$,
because the curvature of the Fermi surface 
around the cold spot decreases as $T$ decreases
due to the growth of the AF fluctuations.
(see Fig. \ref{fig:FS}.)
Whereas $dR_{\rm H}^0/dT<0$ for $T<0.05$
because of the rapid increase of the anisotropy of 
$\gamma_\k$ at lower temperatures.
However, this increase of $R_{\rm H}^0$ is too small to 
explain experimental results.

Finally, we show 
cot$\theta_{\rm H}= \s_{xx}/{\mit\Delta}\s_{xy}$ 
in Fig. \ref{fig:Result} (c).
All the $\cot\theta_{\rm H}$'s are approximately proportional to
$T^2$ below $T\simle 0.05$,
where both $R_{\rm H}\propto T^{-1}$ and $\rho\propto T$ 
are satisfied approximately.
This result is highly consistent with experiments
reported in Ref.
 \cite{Sushko}.
Similar behavior of $\cot\theta_{\rm H}$ is also observed in 
high-$T_{\rm c}$ cuprates.
For high-$T_{\rm c}$ cuprates, Anderson claimed that it suggests 
the non-Fermi liquid ground state
which possesses two kinds of relaxation rates
 \cite{Anderson}.
However, we stress that the relation 
$\cot\theta_{\rm H} \propto T^2$ is naturally understood
both for $\kappa$-(BEDT-TTF)$_2$X salts and for high-$T_{\rm c}$
cuprates within the framework of the nearly AF Fermi liquid.

Finally, 
we discuss the $U$-dependence of $\rho$ and $R_{\rm H}$
in the present calculations,
and compare them with the experimental pressure dependences.
According to the experiments on X=Cu[N(CN)$_2$]Cl
 \cite{Sushko},
as the applied pressure increases, 
(i) $\rho$ decreases monotonously,
(ii) $R_{\rm H}$ is almost unchanged, and
(iii) cot$\theta_{\rm H}$ decreases monotonously.
The effect of the applied pressure is to increase $W_{\rm b}$
while $U$ is unchanged, i.e., to decrease $U/W_{\rm b}$.
Moreover, on condition that $U/W_{\rm b}$ is constant, 
it is easy to see that 
$\rho\propto W_{\rm b}^{-1}$, $R_{\rm H}\propto W_{\rm b}^{0}$,
and cot$(\Theta_{\rm H})\propto W_{\rm b}^{-1}$.
As a result, 
the obtained $U$-dependence of $\rho$ and cot$\theta_{\rm H}$,
which are shown in Fig. \ref{fig:Result} (a) and (c),
are consistent with experiments.
On the other hand, the observed weak pressure dependence of 
$R_{\rm H}$ will correspond to the behavior for $U\ge 8$
in Fig. \ref{fig:Result} (b).
In conclusion, experimentally observed pressure effects
(i)-(iii) are reproduced in our study.

At last, we comment on the superconductivity:
We find $T_{\rm c}=0.024$ for $U=9$ and 
$T_{\rm c}=0.004$ for $U=5$ by solving the Eliashberg equations.
This is consistent with the decrease of $T_{\rm c}$ 
under pressure observed experimentally.
Roughly speaking, $T=0.01$ corresponds to 10K
because the band-width $W_{\rm band}\sim0.5$eV at ambient pressure.
Although the obtained $T_{\rm c}$'s is rather higher than 
experimental one's, it decreases if we put $t'/t$ larger than 0.7
 \cite{Kino}.

\subsection{Derivation of $\s_{\rm inc}$}
In this subsection, we give the derivation of the second term of
eq. (\ref{eqn:s_numerical}) based on the Fermi liquid theory.
We call it the incoherent part of the conductivity $\s_{\rm inc}$
because it gives negligible contribution when the 
quasiparticles are well-defined.
We call the first term of eq. (\ref{eqn:s_numerical})
the coherent part of the conductivity $\s_{\rm coh}$,
which was derived by Eliashberg under the assumption 
$\gamma_\k^\ast \ll T$
 \cite{Eliashberg}.

\begin{figure}
\begin{center}
\epsfig{file=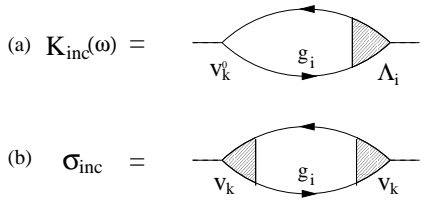,width=5.5cm}
\end{center}
\caption{The diagrams for (a) $K_{\rm inc}(\w)$, and for
 (b) $\s_{\rm inc}$, respectively.
}
\label{fig:Sinc-diagram}
\end{figure}

According to the Kubo formula, the conductivity 
is expressed by the retarded two-particle Green function 
$K^{\rm R}(\w)$,
whose explicit form within the Fermi liquid theory is given by 
eq.(9) of Ref.
 \cite{Eliashberg}.
The first term of eq. (\ref{eqn:s_numerical}),
which was derived by Eliashberg,
comes from the coherent terms of $K^{\rm R}(\w)$
which include at least one $g_2=G^{\rm R} G^{\rm A}$.
Here, we study the contribution from the the 
incoherent part without $g_2$, $K_{\rm inc}^{\rm R}(\w)$,
which has not been analyzed previously.
Hereafter, we omit the momentum variables
for simplicity.
$K_{\rm inc}^{\rm R}(\w)$
is given as follows:
%
\begin{eqnarray}
& &K_{\rm inc}^{\rm R}(\w)
 = -\sum_{i=1,3} \int \frac{d\e}{4\pi\i} v_{\k x}^0 \lambda_i(\e;\w) 
g_i(\e;\w)\Lambda_x^i(\e;\w) ,
 \label{eqn:C1} \\
& &\Lambda_x^i(\e;\w) 
 = v_{\k x}^0 + \sum_{j=1,3}\int \frac{d\e'}{4\pi\i}
  {\cal T}_{i,j}(\e,\e';\w)
 \nonumber \\
& &\ \ \ \ \ \ \ \ \ \ \ \ \ \ \ \ \times
 g_j(\e';\w)\Lambda_x^j(\e';\w) ,
 \label{eqn:C2}
\end{eqnarray}
which is shown in Fig. \ref{fig:Sinc-diagram}(a).
Here, 
$g_1(\e;\w)= G_\k(\e+\w+\i\delta)G_\k(\e+\i\delta)$,
$g_3(\e;\w)= \{ g_1(\e;\w) \}^\ast$, 
$\lambda_1(\e;\w)= {\rm th}(\e/2T)$,
$\lambda_3(\e;\w)= -{\rm th}((\e+\w)/2T)$, and
$v_{\k x}^0 = d\e_\k^0/dk_x$, respectively.
The quantities ${\cal T}_{i,j}(\e,\e';\w)$ are given by the 
product of irreducible four-point vertices
$\Gamma_{i,j}^{\I(\II)}(\e,\e';\w)$ and 
thermal factors brought by the analytic continuation,
whose definition is given by eq. (12) of Ref.
 \cite{Eliashberg}.
(Hereafter, all the four-point vertices are 'irreducible'
with respect to $g_i$.)
We study the conductivity from the incoherent terms,
given by
$\s_{\rm inc}= e^2 \lim_{\w\rightarrow0}
 {\rm Im}K_{\rm inc}^{\rm R}(\w)/\w$,
and find that it gives a finite contribution when the 
life-time of the quasiparticles becomes shorter.

Now, we derive the $\w$-linear term of Im$K_{\rm inc}^{\rm R}(\w)$.
${\cal T}_{i,3}(\e,\e';\w)$ $(i=1,3)$ contains the 
thermal factors ${\rm th}((\e+\w)/2T)$ as shown in eq. (12) of Ref.
 \cite{Eliashberg}.
We can check that $K_{\rm inc}^{\rm R}(\w)$ given by 
eqs. (\ref{eqn:C1}) and (\ref{eqn:C2})
becomes real quantity when we put $\w=0$ (i) in all the factors
${\rm th}((\e+\w)/2T)$ 
and (ii) in all the irreducible vertices $\Gamma_{i,j}(\e,\e';\w)$,
by taking account of the relation
$\Gamma_{i,j}^{\I(\II)}(\e,\e';0)=
 \{ \Gamma_{4-i,4-j}^{\I(\II)}(\e,\e';0) \}^\ast$ for $i,j=1,3$.
This means that the $\w$-linear term of Im$K_{\rm inc}^{\rm R}(\w)$
comes only from one of (i) or (ii).
Because $\Lambda_x^i(\e;\w)$ contains infinite number of
${\cal T}_{i,j}$ $(i,j=1,3)$, we find that 
\begin{eqnarray}
\s_{\rm inc}&=& -e^2\int_{-\infty}^{\infty} \frac{d\e}{\pi} 
 \left(-\frac{\d f}{\d\e} \right) {\rm Re}\left\{ 
 {\tilde \Lambda}_x^1(\e;0)g_1(\e;0)\Lambda_x^1(\e;0) \right\}
 \nonumber \\
&-&e^2\int\!\!\int_{-\infty}^{\infty} \frac{d\e d\e'}{4\pi^2} 
 \sum_{i,j=1,3} \lambda_i(\e;0)
 {\tilde \Lambda}_x^i(\e;0)g_i(\e;0)
  \nonumber \\
& &\times 
 {\bar {\cal T}}_{i,j}(\e,\e') g_j(\e';0) \Lambda_x^j(\e';0)
\label{eqn:C4}
\end{eqnarray}
where ${\bar {\cal T}}_{i,j}(\e,\e')$ is defined as the 
$\i\w$-derivation of the vertex part ($\Gamma_{i,j}$) of 
${\cal T}_{i,j}(\e,\e';\w)$, and put $\w=0$.

As for the second term of (\ref{eqn:C4}),
${\bar {\cal T}}_{i,j}(\e,\e')$ given by the MT-term 
vanishes identically.
On the other hand, the two AL-terms causes the 
finite contribution for Im$K_{\rm inc}^R(\w)$.
After the long calculation, 
${\bar {\cal T}}_{1,1}(\e,\e')$ in the FLEX approximation
is given by 
\begin{eqnarray}
& &{\bar {\cal T}}_{1,1}(\k\e,\k'\e')
 = {\bar \Gamma}_{1,1}^{\I}(\k\e,\k'\e')\cdot {\rm th}(\e'/2T),
 \\
& &{\bar \Gamma}_{1,1}^{\I}(\k\e,\k'\e')
 = -\sum_{\q} {\cal P}\!\int_{-\infty}^\infty\!\frac{dz}{2\pi} 
 \frac{\d {\rm cth}(z/2T)}{\d z} \pi^2
  \nonumber \\
& &\ \ \ \ \ \ \times
 \rho_{\q}(\e'+z) 
 \left( \rho_{\k-\k'+\q}(\e+z)- \rho_{\k-\k'+\q}(\e-z) \right)
  \nonumber \\
& &\ \ \ \ \ \ \times (3U^4/2)\cdot 
 \left[ {\rm Im}\chi_{\k-\k'}^{s}(-z+\i\delta)\right]^2,
  \label{eqn:AL}
\end{eqnarray}
where we have used the relation $v_\k=-v_{-\k}$ in derivation, 
and $\rho_\k(\w)=-{\rm Im}G_\k(\w+\i\delta)/\pi$.
Other terms which do not contribute to $\s_{\rm inc}$
are dropped in eq. (\ref{eqn:AL}).
In the same way of deriving eq. (\ref{eqn:AL}),
we can show that
${\bar {\cal T}}_{i,j}(\k\e,\k'\e')
 =(-1)^{(j-1)/2}\cdot{\bar {\cal T}}_{1,1}(\k\e,\k'\e')$
for $i,j=1,3$.

Now, we show that the second term of eq. (\ref{eqn:C4}) 
is negligible:
It vanishes at $T=0$ because 
Equation (\ref{eqn:AL}) vanishes in this case.
(Note that ${\rm Im}\chi_{\k}^{s}(0)=0$.)
It should be much small even at finite temperatures
because of the cancellation caused by the factor
$(\rho(\e+z)-\rho(\e-z))$
in eq. (\ref{eqn:AL}).
By this reason, we consider only the first term of 
eq. (\ref{eqn:C4}) hereafter.

As for ${\tilde \Lambda}_x$ in eq. (\ref{eqn:C4}),
we can show that
\begin{eqnarray}
& &{\rm Re}\{{\tilde \Lambda}_x^1(0;0) - \Lambda_x^1(0;0)\}= 
 \nonumber \\
& &\ \ \ \ \ \
  \frac12 \sum_{i,j=1,3} \int\frac{d\e}{4\pi i}
 \left[\  {\rm cth}\frac{\e}{2T} -{\rm th}\frac{\e}{2T} \ \right]
 (-1)^{(j-1)/2} 
 \nonumber \\
& &\ \ \ \ \ \ \times 
 \left(\ \Gamma_{i,j}^{\II}(0,\e;0)-\Gamma_{i,j}^{\I}(0,\e;0) \ \right)
 g_j(\e) v_{\k x}^0, 
 \label{eqn:difL}
\end{eqnarray}
which is proportional to $T^2$ at low temperatures.
(Here, $\Gamma_{i,j}^{\I,\II}$ is reducible with respect to $g_i$.)
Thus, ${\tilde \Lambda}_x^1(0;0)$ approximately identical
to $\Lambda_x^1(0;0)$.
We have also estimated eq. (\ref{eqn:difL}) numerically by 
taking only the MT terms into account, and find that the difference
between them is at most a few percent.

In conclusion, we get the following expression for $\s_{\rm inc}$:
\begin{eqnarray}
\s_{\rm inc}
&=& -e^2 \sum_\k \int_{-\infty}^{\infty} \frac{d\e}{\pi} 
 \left(-\frac{\d f}{\d\e} \right) 
 \nonumber \\
& &\ \ \ \ \ \ \ \ \ \ \ \times
 {\rm Re} \! \left\{ \ G_\k^2(\e+\i\delta) \cdot 
 v_{\k x}^2(\e+\i\delta) \ \right\} ,
\label{eqn:C3}
\end{eqnarray}
which is shown in Fig. \ref{fig:Sinc-diagram}(b).
Here we have used the Ward identity
$\Lambda_x^1 (\e;0) 
 = v_{\k x}^0 + \frac{\d}{\d k_x}\Sigma_\k(\e+\i\delta)
  \equiv v_{\k x}(\e+\i\delta)$.
Note that the vertex correction, given by the momentum derivative 
of $\Sigma_\k(\w)$ appears twice in eq. (\ref{eqn:C3}).
Thus the obtained $\s_{\rm inc}$ gives the second term of
eq. (\ref{eqn:s_numerical}).

If we can put
$G_\k(\e)= z_\k/(\e+\mu-\e_\k^\ast +\i\gamma_\k^\ast)$,
eq. (\ref{eqn:C3}) becomes
\begin{eqnarray}
\s_{\rm inc}= 
 e^2 \sum_\k \int \! \frac{d\e}{\pi} 
 \left(-\frac{\d f}{\d\e} \right) 
 z_\k^2 \frac{-(\e-\e_\k^\ast)^2 + {\gamma_\k^\ast}^2}
  {( (\e-\e_\k^\ast)^2 + {\gamma_\k^\ast}^2 )^2} v_{\k x}^2,
\label{eqn:inc-est}
\end{eqnarray}
where $\gamma_\k^\ast= z_\k\gamma_\k$ and $\e_\k^\ast= z_\k\e_\k$,
respectively.
In the case of $\gamma_\k^\ast\ll T\ll W_{\rm b}$,
$\s_{\rm inc}\approx 0$ is realized
according to eq.(\ref{eqn:inc-est}),
while $\s_{\rm coh}\propto \gamma_{\rm cold}^{-1}$.
However, such a condition is not satisfied in the present 
calculation as shown in Fig. \ref{fig:J}(a).
At lower temperatures, $\gamma_\k^\ast \simle T$ is satisfied
because $z_\k^{-1} \simle 10$ then, whereas $\gamma_\k^\ast \simge T$
at higher temperatures because $z_\k^{-1}$ decreases
as $T$ increases.
Thus, $\s_{\rm inc}$ is expected to be important at 
higher temperatures.
In Appendix A, we show its importance numerically.

\section{Comparison with Experiments}
 
\subsection{Effect of Band-Splitting around the Hot-Spots}
In this section, we compare our theoretical results with
experiments in more detail.
Here, we discuss the validity of the present results
based on the effective model shown in Fig. \ref{fig:FS}.
Precisely speaking,
in many (not all) real $\kappa$-(BEDT-TTF)$_2$X compounds,
the Fermi surface splits slightly around the hot spots 
in Fig. \ref{fig:FS} because a unit cell contains two
dimers of the BEDT-TTF molecules 
 (see Appendix B).
In this sense, the present model may be too simplified for the
quantitative studies.

However, the mechanism of the enhancement of $R_{\rm H}$ 
due to the AF fluctuations proposed in this paper
is surely valid because  
only the quasiparticles around the cold spots plays an important 
role for transport phenomena 
as shown in Fig. \ref{fig:J}.
On the other hand, $R_{\rm H}$ at higher temperatures
may be affected by the splitting of the Fermi surface
at the hot spots.
Thus, our calculation based on the dispersion, eq. (\ref{eqn:disp1}),
is comparable with experiments
at least in the lower temperature region.


\subsection{Effect of Temperature Dependence of the Volume}
Next, we discuss the effect of the thermal contraction of 
the volume, which is known to be quite large in various organic metals.
 \cite{Nogami,Jerome}.
For example, (TMTSF)$_2$PF$_6$ at ambient pressure shows 
$\rho\propto T^n$ and $n\approx2$,
whereas $n\approx1$ is concluded after the 
effect of the thermal contraction is compensated
 \cite{Jerome}.
 
As for X=Cu[N(CN)$_2$]Cl,
$\rho\propto T^n$ and $n\approx2$
is observed below 100K at nearly ambient pressures
 \cite{Kanoda-review,Dressel}.
However, $n\approx1$ is realized qualitatively
under the constant-volume condition according to Ref.
 \cite{Sushko}:
In the article, 
the authors used the piston-cylinder clamped cell
to make pressure,
and the oil inside of the cell freezes at $\sim$200K,
which make the volume of the sample constant
 \cite{private}.
This experimental fact means that 
$n\approx2$ at ambient pressure should not be interpreted as 
the conventional Fermi liquid behavior.
Note that
the non-Fermi liquid behavior $n\approx1$ is observed
under pressure, nonetheless $U/W_{\rm b}$ decreases by pressure
 \cite{comment}.

Thus, the results obtained in our study,
$\rho\propto T$ and $R_{\rm H}\propto T^{-1}$,
are qualitatively consistent with 
Ref.\cite{Sushko},
which are expected to be the intrinsic behaviors in $\kappa$-BEDT-TTF
compounds without the volume contraction.
We stress that,
as for organic compounds,
the measurements under constant volume condition are highly
demanded for the comparison with theories.

\subsection{The Saturation of $R_{\rm H}$ at Lower Temperatures}
Here, we comment on the saturation of 
$R_{\rm H}$ below a characteristic temperature $T^\ast$ 
observed experimentally.
Around $T\approx T^\ast$, the $1/T_1T$ also saturates and begins to 
decrease below $T^\ast$, which is called the pseudo spin-gap behavior.
This results suggests that $\xi_{\rm AF}^2$, or $\chi_Q^s(0)$,
will saturates below $T^\ast$.
(We note that many experiments for $1/T_1T$ are done
at ambient pressure, so the volume contraction effect may 
play some quantitative effect on $1/T_1T$ at low temperatures.)
Thus, the analytical relation in our work,
$R_{\rm H} \propto \xi_{\rm AF}^2$,
is consistent with these experiments.

However, $R_{\rm H}$ of our numerical calculation 
in Figs. \ref{fig:Result} does not saturate:
This is because the FLEX approximation does not reproduce the 
saturation of $\xi_{\rm AF}^2$ below $T^\ast$,
which is a significant future problem.
One of the possible mechanism for it will be the precursor effect
of superconductivity below $T^\ast$
 \cite{Jujo}.
Finally, we point out that
$R_{\rm H}$ begins to decrease on cooling below $T^\ast$ 
in under-doped high-$T_{\rm c}$ cuprates,
whereas it takes a saturate value at lower temperatures
in X=Cu[N(CN)$_2$]Cl
 \cite{Sushko}.

\section{Conclusions}
In this paper, we have presented the theoretical study for the 
resistivity and the Hall coefficient of the $\kappa$-(BEDT-TTF)$_2$X 
salts.
Reference
 \cite{Sushko}
point out some experimental evidences that 
there anomalous behaviors have close connection
with the grows of the AF fluctuations 
 \cite{Comment-MR}.
According to our theory,
the Hall coefficient follows the relation
$R_{\rm H}\propto \xi_{\rm AF}^2 \propto T^{-1}$
in nearly AF Fermi liquid state,
which is consistent with the experiment
under the constant volume condition
 \cite{Sushko}.
This anomaly of $R_{\rm H}$,
which can not be reproduced by the RTA,
is found to come from the VC's for the current
which is indispensable to satisfy the conserving laws.

Moreover, based on the Kubo formula, 
we have derived the expression of the incoherent conductivity 
$\s_{\rm inc}$ beyond the Eliashberg's transport theory,
and found that it give a qualitatively important contribution
in $\kappa$-(BEDT-TTF)$_2$X and in high-T$_{\rm c}$ cuprates
at higher temperatures, where $\gamma_\k^\ast \sim T$ is realized.
It also give the appropriate temperature dependence of $R_{\rm H}$.

We have calculated both $\rho$ and $R_{\rm H}$
based on the FLEX approximation for $U=6\sim10$,
without assuming any fitting parameters.
The obtained $U$-dependences for $\rho$, $R_{\rm H}$ 
and cot$\theta_{\rm H}$
explain well the experimentally observed pressure dependences.
In conclusion, many essential electronic properties of 
$\kappa$-(BEDT-TTF)$_2$X,
expecially both the anomalies of transport phenomena
and the phase diagram,
are explained well from the standpoint of the 
nearly AF Fermi liquid state.
We stress that 
further observations under the constant volume condition
are highly demanded for organic metals
to make a meaningful comparison between theory.

\acknowledgements
We are grateful to T. Moriya, K. Yamada, K. Ueda, and K. Kanki
for valuable comments and discussions.
We also thank Murata and Y. Nogami for the important
information on experiments.

\appendix
\section{Importance of the Incoherent Conductivity}
\begin{figure}
\vspace{10mm}
\begin{center}
\epsfig{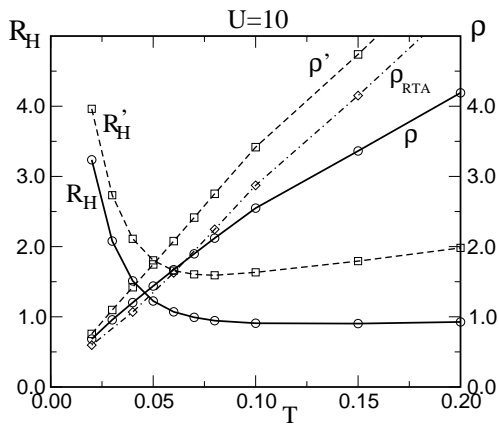}
\end{center}
\vspace{10mm}
\caption{(i) Comparison between $\rho=1/(\s_{\rm coh}+\s_{\rm inc})$ 
and $\rho'=1/\s_{\rm coh}$ for $U=10$.
$\rho$ becomes smaller than $\rho'$
because of the incoherent part of the resistivity.
$\rho_{\rm RTA}$ is given by the RTA.
(ii)Comparison between 
$R_{\rm H}=({\mit\Delta}\s_{xy}/B)\cdot \rho^2$ and
$R_{\rm H}'=({\mit\Delta}\s_{xy}/B)\cdot \rho'^2$ for $U=10$.
We see that $dR_{\rm H}'/dT$ becomes positive below
$T\approx0.08$, which is inconsistent with 
experiments.
}
\label{fig:Cohinc}
\end{figure}

In this appendix, we numerically show the important role 
of $\s_{\rm inc}$ to get the reasonable behaviors 
of $\rho$ and $R_{\rm H}$.
Figure \ref{fig:Cohinc} shows the temperature dependence of
the Hall coefficient and the resistivity.
$\rho=1/(\s_{\rm coh}+\s_{\rm inc})$, $\rho'=1/\s_{\rm coh}$ and 
$\rho_{\rm RTA}=1/\s_{\rm RTA}$, respectively.
Here, $\s_{\rm RTA}$ is given by replacing $J_\k$ with $v_\k$  
in $\s_{\rm coh}$, which is equal to the result from 
the relaxation time approximation.
We find that
(i) $\rho'>\rho_{\rm RTA}$ because of the VC's for the current,
and (ii) $\rho<\rho'$ because of $\s_{\rm inc}$,
which becomes dominant especially at higher temperatures.
As a result, $\rho < \rho_{\rm RTA}$ is realized at higher
temperatures.


As for Hall effect,
$R_{\rm H}= ({\mit\Delta}\s_{xy}/B)\cdot\rho^2$ and
$R_{\rm H}'= ({\mit\Delta}\s_{xy}/B)\cdot{\rho'}^2$, respectively.
We see that $R_{\rm H}<R_{\rm H}'$ because $\rho<\rho'$.
However, $R_{\rm H}(T=0.02)/R_{\rm H}(T=0.2) $ is larger 
than that of $R_{\rm H}'$, so the incoherent conductivity
make the temperature dependence of $R_{\rm H}$ larger.
We stress that $dR_{\rm H}'/dT$ become positive at higher temperature,
which contradicts with experiments
 \cite{Sushko}.
In conclusion, we find that $\s_{\rm inc}$ is necessary to reproduce 
the reasonable behavior of $R_{\rm H}$.


\section{The more precise tight-binding model for 
 $\kappa$-(BEDT-TTF)$_2$X} 
\begin{figure}
\begin{center}
\epsfig{file=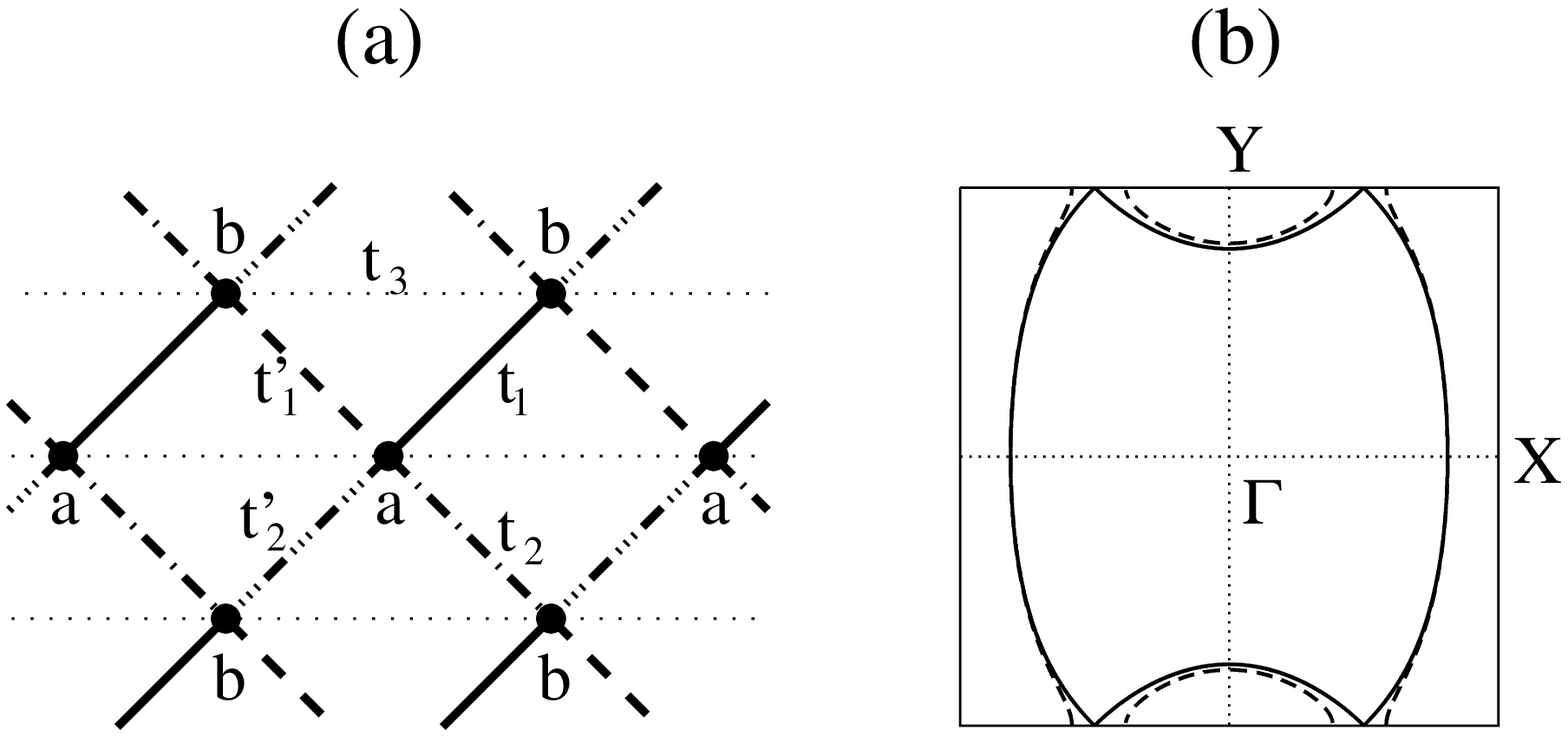,width=8cm}
\end{center}
\caption{}
{\small
(a): The effective model for $\kappa$-(BEDT-TTF)
salts with two-sites in a unit cell ($a,b$).
The hopping parameters of this model 
are given by the parameters in Fig.1 of Ref. 
\cite{HF-approx}
as
$t_1=(p+q')/2$, $t_1'=(p+q)/2$, 
$t_2=(p'+q)/2$, $t_1'=(p'+q')/2$, and
$t_3=b_2/2$, respectively.
(b): The Fermi surface for $t_1=t_1'=t_2=t_2'= 1$
(full line), and for $t_1=t_2'=1.1$ and $t_2=t_1'= 0.9$ 
(broken line), respectively.
In both cases, $t_3= 0.7$.
Note that the former is equivalent to Fig.1
in the extended zone representation.
}
\label{fig:titebinding}
\end{figure}

In real $\kappa$-(BEDT-TTF)$_2$X systems, there are two pairs of 
closely-packed BEDT-TTF molecules in a unit call, 
and only the bonding-orbit of each closely-packed molecules
contributes to make the Fermi surface.
By taking the results of the band-calculations into account,
we get the effective tight-binding model for $\kappa$-(BEDT-TTF)$_2$X
as shown in Fig. \ref{fig:titebinding},
where there are two sites (a,b) in a unit cell.
Each site corresponds to a closely-packed BEDT-TTF molecules.
This model becomes equal to the anisotropic triangular lattice model
given by Fig. \ref{fig:FS} if we put
$t_1=t_1'=t_2=t_2'\equiv t$ and $t_3\equiv t'$.

The dispersion of the tight-binding model
given by Fig. \ref{fig:titebinding} for $U=0$ is derived as
$\e_\k^\pm= 2t_3\cos k_y \pm [ t_1^2+{t_1'}^2+t_2^2+{t_2'}^2
 +2(t_1t_1'+ t_2t_2')\cos k_y + 2(t_1't_2+t_1t_2')\cos k_x 
 +2t_1t_2'\cos(k_x+k_y) +2t_1't_2\cos(k_x-k_y) ]^{1/2}$.
When $|k_y|=\pi$, 
then $\e_\k^+-\e_\k^- = 2[ (t_1-t_1')^2+(t_2-t_2')^2 ]^{1/2}$.
This means that the Fermi surface splits
around the hot spots when $t_1\ne t_1'$ or $t_2\ne t_2'$,
which is realized in many systems.
(However, both $t_1=t_1'$ and $t_2=t_2'$ are satisfied exactly
in some compounds exceptionally, e.g.,
 \cite{CN3}.)
This splittings of the Fermi surface at the hot spots 
will not affect the temperature dependence of $R_{\rm H}$
at low temperatures, as discussed in \S III.



\end{multicols}

\begin{thebibliography}{99}
\bibitem{SDW-comment}
According to the analysis,
${\vec J}_{\k}$ becomes parallel to the magnetic Brillouin zone
at the hot spot when $\xi_{\rm AF} \rightarrow \infty$,
which may be interpreted as the precursor effect of 
the AF ordered state.
\bibitem{comment}
 When $\rho\propto T^2$ is observed in 
 $\kappa$-(BEDT-TTF)$_2$X compounds, in general,
 (i) the volume contraction is prominent, and 
 (ii) $\rho$ shows an insulating behavior at higher temperatures.
\bibitem{Comment-MR} 
 The experimentally observed 'positive magnetoresistance' 
 is also reproduced by the VC's theoretically
 if the AF fluctuations are dominant 
 [H. Kontani: unpublished].
\end{thebibliography}
\end{document}